\documentclass[twocolumn]{aastex631}
\usepackage{graphicx}

\def\simgr{\,\hbox{\hbox{$ > $}\kern -0.8em \lower 1.0ex\hbox{$\sim$}}\,}
\def\simle{\,\hbox{\hbox{$ < $}\kern -0.8em \lower 1.0ex\hbox{$\sim$}}\,}

\newcommand{\xmm}{XMM-Newton}
\newcommand{\swj}{Swift~J0503.7$-$2819}

\def\rxj{RX~J0838.7$-$2827}
\def\pola{1RXS~J083842.1$-$282723}
\def\polb{IGR~J19552+0043}

\def\cpd{cycles~day$^{-1}$}

\shortauthors{Halpern}
\shorttitle{Periods of the Asynchronous Polar \rxj}

\begin{document}
\title{Resolving the Periods of the Asynchronous Polar \pola}

\author[0000-0003-4814-2377]{J. P. Halpern}
\affiliation{Department of Astronomy and Columbia Astrophysics Laboratory, Columbia University, 550 West 120th Street, New York, NY 10027-6601, USA; jph1@columbia.edu}

\begin{abstract}

  \pola\ is a nearly synchronous magnetic cataclysmic variable with a simple X-ray light curve. While its orbital period was fairly well established at $P_{\rm orb}=98.4$ minutes from optical spectroscopy, indirect estimates of $P_{\rm spin}/P_{\rm orb}$ ranged from 0.90 to 0.96 because the short X-ray light curves could not determine the beat period to a factor of 2.  We analyze a recent 50 day TESS observation, and ground-based optical time-series photometry spanning 9 years, that together measure precise beat, orbit, and spin periods and enable the X-ray and optical modulations to be phase aligned.  Although the X-ray light curves do not distinguish between a beat period of 16.11 or 32.22 hours, all of the optical evidence favors the longer value, with complete pole switching of accretion every half beat cycle.  This would require $P_{\rm spin}/P_{\rm orb}=0.952$. Long-term optical monitoring also shows a decline in accretion rate, and a change in the beat-folded light curve.  It would be useful to obtain a new X-ray/optical observation of at least 32~hours duration to examine any associated change in accretion structure, and confirm the spin and beat periods.

\end{abstract}

\section{Introduction\label{sec:intro}}

Cataclysmic variables (CVs) are accreting binaries in which a dwarf
star donates mass to a white dwarf (WD) via Roche-lobe overflow.
A strong WD magnetic field can truncate an accretion
disk at the magnetospheric boundary, or even prevent a disk from forming.
In polars (AM Her stars), the magnetic field is strong enough to
channel an accretion stream directly from the donor onto a magnetic pole,
and lock the WD rotation to the binary orbit ($P_{\rm spin} = P_{\rm orb}$).
Intermediate polars (IPs) have weaker $B$-fields; the WD poles(s) can be
fed either from a truncated disk, a diskless stream, or a combination
of both (disk-overflow stream).
The spin of the WD in an IP is detected as a coherent oscillation
in X-ray or optical emission at a period shorter than the
orbital period of the binary, typically with
$P_{\rm spin} \le0.1\,P_{\rm orb}$.

A group of four ``asynchronous polars'' (APs), have spin
and orbital periods that differ by $<2\%$.  One of these, V1500 Cyg, had a nova
explosion in 1975, and because it is generally observed that APs
are evolving toward synchronism on times scales of 100--13,000 yr
\citep{mye17,lit23a} it is plausible that nova eruptions perturbed the spin of the WD.  A fifth member of this class is
SDSS J085414.0+390537 \citep{kol23}.

Several magnetic CVs have been discovered with
greater degree of asynchronism than the APs.
IGR J19552+0044 \citep{tov17}, SDSS J084617.1+245344, 
SDSS J134441.8+204408 \citep{lit23a,lit23b}, and ``Paloma''
\citep{sch07,jos16,lit23a}
have $P_{\rm spin} = (0.87-0.97)\,\,P_{\rm orb}$.
It is unlikely that nova eruptions could perturb the spin by this much.
But it is not clear whether these objects are
pre-polars approaching synchronism,
or stream-fed IPs whose spins are in stable equilibrium.

Fundamental to the life cycle of an mCV is the change of
its equilibrium spin period as it evolves
to shorter orbital period.
In numerical simulations of accretion and spin equilibrium,
\citet{nor04,nor08} concluded that
$P_{\rm spin}/P_{\rm orb}$ should increase as $P_{\rm orb}$ decreases,
and proceed to synchronism once $P_{\rm spin}/P_{\rm orb}>0.6$.
That most of the nearly synchronous mCVs have short orbital periods
where IPs are rare, below the 2--3 hour period gap,
supports the theory.  Such compact systems are thought to accrete
from a thin ring near the edge of the Roche lobe of the white dwarf.
However, the values of the parameters of the theory, particularly the
stellar mass ratio of 0.5, were chosen when nearly all IPs were
known to have $P_{\rm spin}/P_{\rm orb}<0.6$ (with the exception of EX Hya at 0.68).
This choice had the effect of ``predicting'' that
IPs with $0.6<P_{\rm spin}/P_{\rm orb}<1$ should be rare.
But the number of recent asynchronous discoveries
suggests that they are not so rare, and that spin equilibrium is possible at most if not all values of
$P_{\rm spin}/P_{\rm orb}$.  This could occur if the mass ratio is $<0.5$,
as it would be at small orbital periods, and also if the
magnetic moment of the secondary is weak, preventing the system from
synchronizing \citep{nor04}.

Two more members of this family, \swj\ \citep{hal22,raw22,pra23}
and \pola\ \citep[hereafter Paper~1]{rea17,hal17} have uncertain
degrees of asynchronism because X-ray coverage of their spin-orbit beat
periods was not complete.   The favored interpretation of X-ray
and optical data on \swj\ yields $P_{\rm spin} = 0.803\,\,P_{\rm orb}$.
The subject of this paper is a proposed solution for \pola\
(hereafter \rxj) enabled by TESS, the the Transiting Exoplanet Survey Satellite
\citep{ric15}.

The position of \rxj\ is (Gaia-CRF3, epoch 2016.0) R.A.=$08^{\rm h}38^{\rm m}43^{\rm s}\!.3315$, decl.=$-28^{\circ}27^{\prime}00^{\prime\prime}\!.944$ \citep{gai16,gai23}.  Its parallax is $6.3855\pm0.0828$~mas, corresponding to a distance of 157~pc.  Proper motion components are ($\mu_{\alpha}{\rm cos}\,\delta, \mu_{\delta})=(-16.93\pm0.07, +12.46\pm0.08)$~mas~yr$^{-1}$.  Its average X-ray luminosity is $\approx2\times10^{31}$ erg~s$^{-1}$ in the 0.3--10~keV band from \xmm\ (Paper~1).

Section~\ref{sec:obs} reviews the X-ray light curves of \rxj\ and their original interpretations, followed by analysis of the recent TESS observation and long-term ground-based optical timing of \rxj\ that refine
the periods to high precision.  Section~\ref{sec:disc} further evaluates the alternative identifications of the
beat and spin periods.  Conclusions and suggestions for further study of this variable object are presented in Section~\ref{sec:conc}.

\begin{figure*}
\vbox{\hspace{0.6in}
\includegraphics[angle=0.,width=0.8\linewidth]{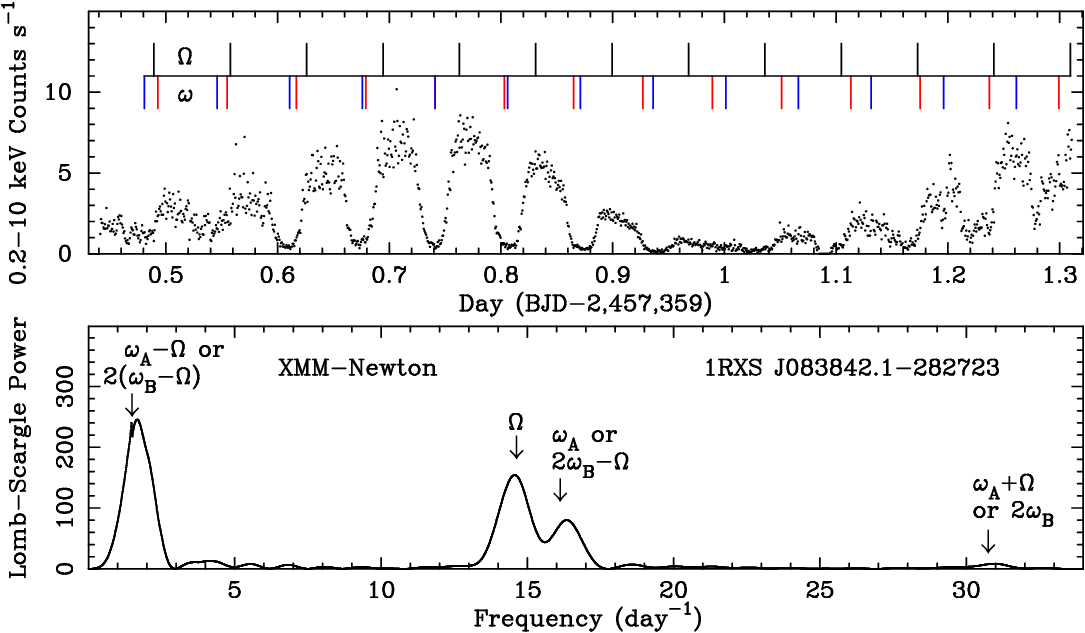}
}
\vspace{0.1in}
\vbox{\hspace{0.6in}
\hspace{-0.13in}
\includegraphics[angle=0.,width=0.8124\linewidth]{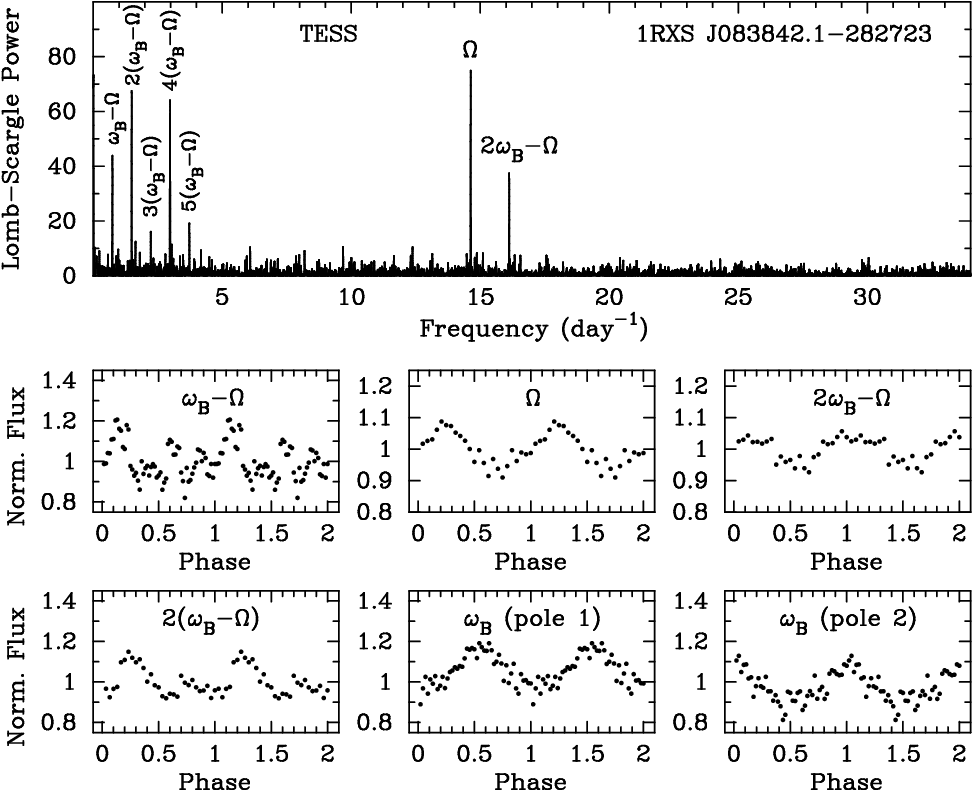}
}
\caption{
Top: From the \xmm\ observation of \rxj\ on 2015 December 2,
a combined EPIC pn and MOS 0.2--10~keV light curve in 60~s bins.
There is a short gap in the data at day 1.09.  The alternative
spin cycles are marked in red (Case~A: short beat cycle) and blue
(Case~B: long beat cycle), where blue corresponds to pole
switching at each minimum of the beat cycle.
Middle: Periodograms of the \xmm\ light curve and 50 day TESS observation.
Peaks in \xmm\ are identified in either Case~A (upper label)
or Case~B (lower label), while TESS favors Case~B.  
Bottom: TESS light curve folded on the Case~B beat, orbital, and spin periods,
revealing different fluxes from accreting poles 1~and~2.
}
\label{fig:rxjfig}
\end{figure*}

\section{Periods from X-ray and Optical\label{sec:obs}}

\subsection{X-ray Light Curves (2015, 2016)\label{sec:xray}}

\rxj\ was observed twice by \xmm\ and once by Chandra (Paper~1);
we review the resulting interpretations here.
The longest of the \xmm\ light curves
is reproduced in Figure~\ref{fig:rxjfig}.  The peaks and dips in the X-ray light
curve suggest self-occultation by the rotating white dwarf
of a stream-fed emitting spot whose accretion rate is strongly modulated
at the longer beat frequency between spin and orbit.
But the observation was not long enough to resolve well the spin and orbit.
\citet{rea17} interpreted the blended pair of peaks around 15~\cpd\ in the
X-ray periodogram of Figure~\ref{fig:rxjfig} 
as nearly synchronous spin ($\omega$) and orbital ($\Omega$) frequencies,
and a peak near 1.6~\cpd\ as the beat frequency, $\Omega_{\rm beat}=\omega-\Omega$, corresponding to a beat
period of $\approx15$~hours.
The beat-modulated intensity could be caused switching of the
accretion to an unseen ``lower'' pole, and back to the
upper pole within one beat cycle.  We denote this scenario ``Case A.''

Noting that the spin period in Case~A does not match
well the actual spacing between the X-ray pulses,
Paper~1 outlined a ``Case B,'' in which the time between adjacent dips
suggests a slightly longer spin period that falls midway
between the blended pair in the periodogram.  Assuming that both poles
are visible, and switching of accretion between them happens
during a minimum of the beat modulation, the beat period is
twice as long as in Case~A.  The second half of the beat cycle, which was not completely covered, is emission from the second pole. In the
event of such complete pole switching, the spin signal $\omega$
is suppressed in the power spectrum, being replaced by the two sidebands
$\omega\pm\Omega_{\rm beat}=\Omega$ and $2\omega-\Omega$ (Wynn \& King 1992).
These two are the Case~B identifications of the split peak.  

In Paper~1 the peak in the periodogram at $\approx1.6$~\cpd\ was interpreted as $3(\omega_{\rm B}-\Omega)$.  This, however, was based on imprecise knowledge of the
spin or beat periods.  The subsequent optical observations discussed below show
that this peak is contributed primarily by $2(\omega_{\rm B}-\Omega)$.  Note that the
values of the frequencies marked in Figure~\ref{fig:rxjfig}
are taken from the precise optical results listed in Tables~\ref{tab:tess}
and~\ref{tab:mdm}.  A spot model created by \citet{wan20} using the results of Paper~1 to match the X-ray light curve and power spectrum 
could be revised using these new values.

The difference between Case~A and Case~B is indicated by the tick
marks above the X-ray light curve. The orbital period is the
same in both cases, and agrees with optical radial velocity
spectroscopy (Paper~1).  The alternative spin periods are
illustrated by the red (Case~A) and blue (Case~B) tick marks,
with Case~B corresponding to complete
pole switching between each half of the beat cycle.  This is exactly 
the dilemma of whether to identify the peak to the right of $\Omega$
in the periodogram as $\omega_{\rm A}$ or $2\omega_{\rm B}-\Omega$.

\subsection{TESS (2023 January 18 -- March 10)\label{sec:tess}}

A 50 day light curve of \rxj\ at 2~minute cadence was obtained by the
Transiting Exoplanet Survey Satellite (TESS; \citealt{ric15}) in its
sectors 61 and 62, starting on 2023 January 18.  The average magnitude derived from the PDCSAP
(Pre-search Data Conditioning Simple Aperture Photometry) fluxes is 18.2, consistent with simultaneous ground-based $r$-band observations presented in Section~\ref{sec:mdm}. We merged the two sectors' light curves
to calculate the periodogram in Figure~\ref{fig:rxjfig}.  It has prominent
narrow peaks, listed in Table~\ref{tab:tess}
corresponding to the ones detected but not clearly resolved by
the short \xmm\ observation.  The strongest signal, at 14.63~\cpd,
is consistent with the orbital frequency $\Omega$ first determined from optical
radial velocity spectroscopy, while a peak at 16.12~\cpd\
is either the spin $\omega_{\rm A}$ or $2\omega_{\rm B}-\Omega$.

There are five equally spaced peaks between 0.745~\cpd\ and 3.726~\cpd\
that can be attributed to the beat frequency $\omega_{\rm B}-\Omega$
and its harmonics.  These may help to resolve
the ambiguous identity of the 16.12~\cpd\ peak.
If 0.745~\cpd\ is $\omega_{\rm B}-\Omega$, then the peak at
16.12~\cpd\ should be $2\omega_{\rm B}-\Omega$.  Otherwise, if 16.12~\cpd\ were
$\omega_{\rm A}$, then 0.745~\cpd\ would have to be ${1\over2}(\omega_{\rm A}-\Omega)$.
This and the higher half-integer harmonics would
not have a natural interpretation.  Thus, the TESS peaks
in Figure~\ref{fig:rxjfig} are labeled according to their
Case~B identifications, in which there are two minima per beat cycle
and complete pole-switching occurs during each minimum.  

The bottom panels in Figure~\ref{fig:rxjfig} show the TESS data folded on
the major periods in the power spectrum.  The epochs of phase zero are defined in Section~\ref{sec:ephem} and Table~\ref{tab:ephem}.  When deciding between Case~A
and Case~B, evidence of non-identical poles would support
Case~B's complete pole switching.  Even
if the spin frequency is not present in the total power spectrum, pulses
from opposite poles can be detected by folding half or less of the Case~B
beat cycle --- in this application we choose 40\% --- on the inferred spin period, and sliding the folding window through the
beat cycle in search of maximum power in the spin modulation.

Using this method
we found two regions, approximately $180^{\circ}$ apart in longitude,
with pulsed fractions of
$\approx10\%$ but differing by $\approx10\%$ in average flux. 
This small but significant difference is what contributes power to and identifies
0.745~\cpd\ as the fundamental beat frequency $\omega_{\rm B}-\Omega$. This, in
turn, requires the 16.12~\cpd\ peak to be $2\omega_{\rm B}-\Omega$ rather than $\omega_{\rm A}$.
Thus, TESS provides independent evidence in favor of Case~B. Note that the beat cycle of frequency $\omega_{\rm B}-\Omega$ is complex in TESS, comprising four peaks, unlike the simpler beat structure in X-rays.  This explains why even multiples of the beat
frequency are strong in TESS, and suggests that the accretion structure changed between 2015 and 2023.

\begin{deluxetable}{llcc}
\label{tab:tess}
\tablecolumns{4} 
\tablewidth{0pt} 
\tablecaption{TESS Observed and Inferred Frequencies}
\tablehead{
 \colhead{Frequency} & \colhead{Period} & \colhead{Case A}
 & \colhead{Case B} \\
 \colhead{(day$^{-1}$)} & \colhead{} & & 
}
\startdata
& \hfill Observed & & \\
\hline
0.7452(22)  &  32.206(95)  hr & ${1\over2}(\omega_{\rm A}-\Omega)$ & $\omega_{\rm B}-\Omega$ \\
1.4918(15)  &  16.088(16)  hr & $\omega_{\rm A}-\Omega$  & $2(\omega_{\rm B}-\Omega)$ \\
2.2317(35)  &  10.754(17)  hr & ${3\over2}(\omega_{\rm A}-\Omega)$ & $3(\omega_{\rm B}-\Omega)$ \\
2.9782(17)  &   8.0586(46) hr & $2(\omega_{\rm A}-\Omega)$ & $4(\omega_{\rm B}-\Omega)$ \\
3.7262(29)  &   6.4409(50) hr & ${5\over2}(\omega_{\rm A}-\Omega)$ & $5(\omega_{\rm B}-\Omega)$ \\
14.6290(15) &  98.435(10) min & $\Omega$            & $\Omega$ \\
16.1202(23) &  89.329(13) min & $\omega_{\rm A}$            & $2\omega_{\rm B}-\Omega$ \\
\hline
&  \hfill Inferred  & & \\
\hline
15.3746(27) &  93.661(16) min  & \dots &  $\omega_{\rm B}$ \\ 
\enddata
\end{deluxetable}

\begin{figure*}[t]
\centerline{
\includegraphics[angle=0.,width=0.8\linewidth]{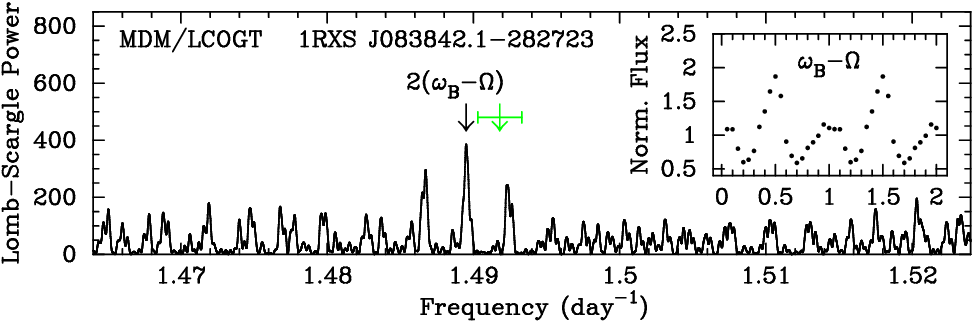}
}
\centerline{
\includegraphics[angle=0.,width=0.8\linewidth]{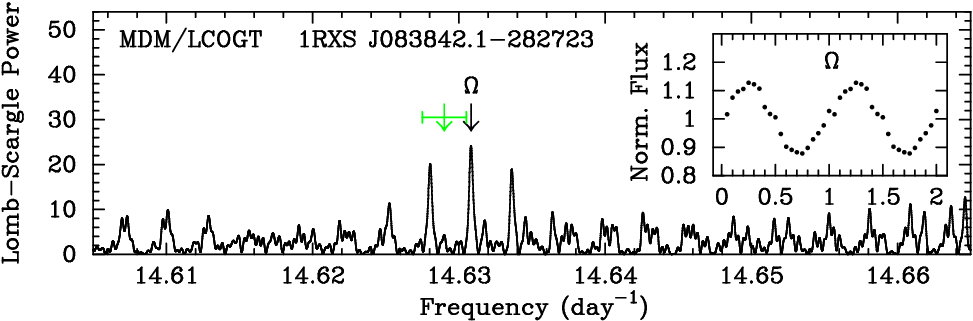}
}
\centerline{
\includegraphics[angle=0.,width=0.8\linewidth]{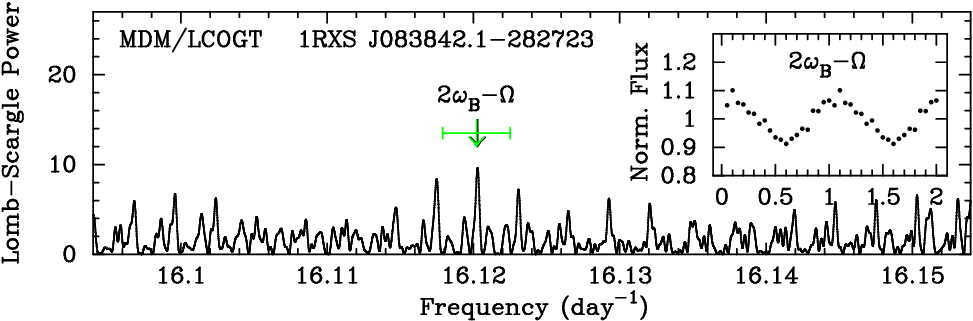}
}
\caption{
Periodograms of MDM and LCOGT data spanning 2014--2023, showing probable detections of 
the main periods also found by TESS and assigned their Case~B identifications.
Green symbols mark the values and uncertainties from TESS.
One-year aliases flank the highest peaks.
Inserts are the light curves folded at the highest MDM peak, except that the beat frequency ($\omega_{\rm B}-\Omega$) is half the frequency of the detected peak in the upper panel.
}
\label{fig:mdm}
\end{figure*}

\begin{figure*}[t]
\centerline{
  \includegraphics[angle=0.,width=1.0\linewidth]{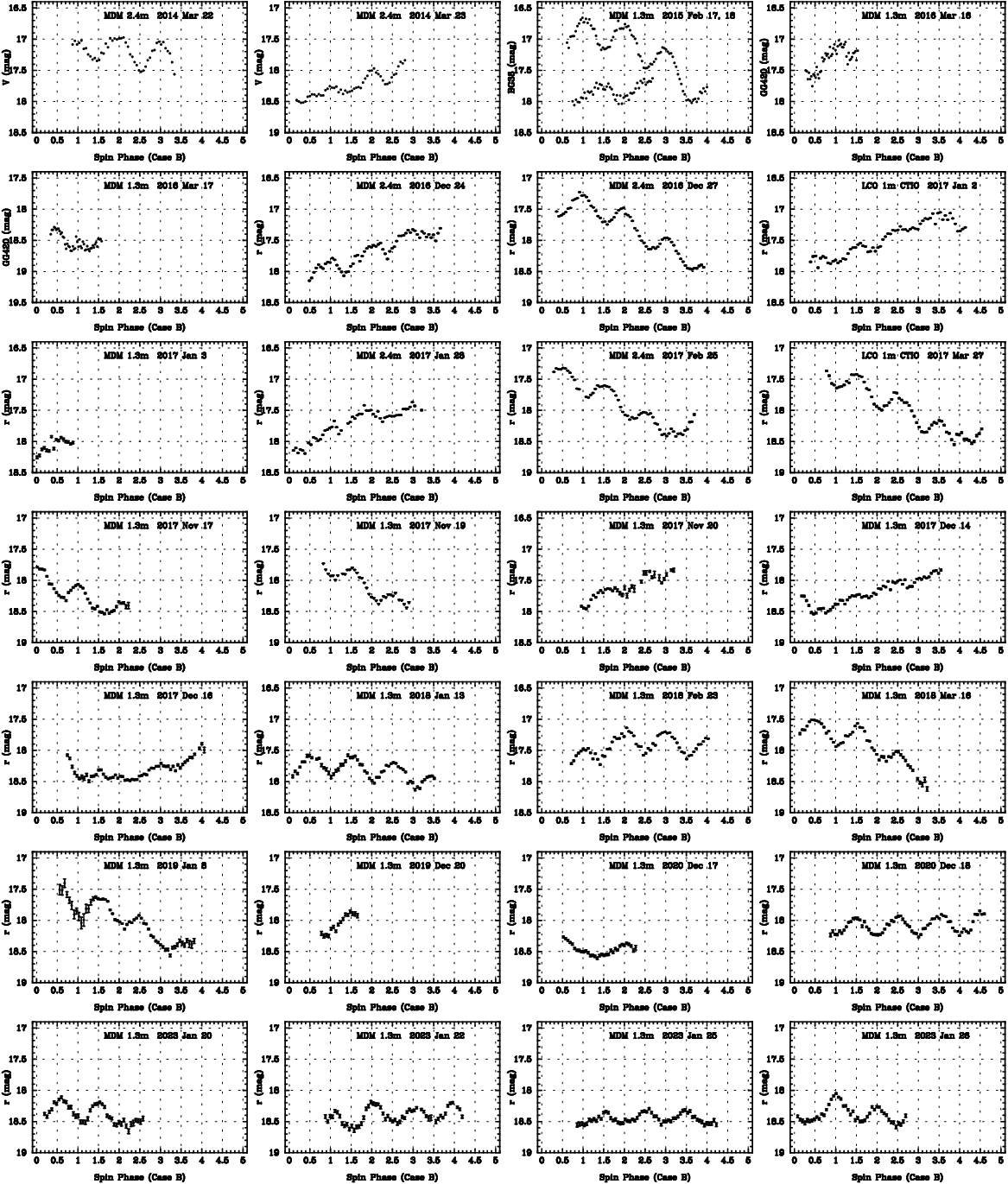}
}
\caption{
MDM and LCOGT light curves from 2014--2023, folded on the Case~B spin period from
Table~\ref{tab:mdm}.  The six earliest
light curves (from Paper~1), which were obtained with cadences of 5--20~s in a $V$, BG38,
or GG420 filter, have been binned to either 150~s or 200~s to better match the signal-to-noise and 5~minute cadence of the subsequent $r$-band data.   
}
\label{fig:phase_b}
\end{figure*}

\subsection{Ground-based Optical (2014--2023)\label{sec:mdm}}

Paper~1 reported a spectroscopic period of 0.068342(3)~day from
strong, single-peaked Balmer and \ion{He}{1} emission lines, which we identified as the orbital period.  The radial velocity amplitude
was $274\pm16$~km~s$^{-1}$. Time-series
photometry collected for 2--7 hours per night on six nights in 2014--2016
in various filters showed oscillations with a period of $\approx0.07$~day
and changes of $\approx1.5$~mag between adjacent nights, corresponding to
modulation at the X-ray beat cycle.  Subsequent to that study, we obtained
time series photometry on 21 nights between 2016 December and 2023 January using
either the 2.4~m or 1.3~m telescope of the MDM Observatory, and on two nights using
a CTIO 1~m telescope of the Las Cumbres Observatory global telescope network
(LCOGT), all in an SDSS $r$ filter with 5~min cadence.

Coherent power-spectrum analysis (Figure~\ref{fig:mdm}) of the combined
29 light curves spanning 9 years shows periods corresponding to the
strongest peaks in the TESS periodogram, but with much higher precision,
as listed in Table~\ref{tab:mdm}.
Although 1-year aliases are also consistent with the TESS periods
at $2(\omega_{\rm B}-\Omega)$, $\Omega$, and $2\omega_{\rm B}-\Omega$,
only the highest peak in each case gives consistent phasing of all
of the optical and X-ray data.  This can be seen by comparing the
folded light curves in the insets of Figure~\ref{fig:mdm}
with the folded TESS light curves in Figure~\ref{fig:rxjfig}.  Note that only $2(\omega_{\rm B}-\Omega)$ is detected in the ground-based periodogram because it has higher power, but we fold on $\omega_{\rm B}-\Omega$ in Figure~\ref{fig:mdm}.

The only major difference between the folds of the TESS and ground-based data
is the modulation at the beat period $\omega_{\rm B}-\Omega$. While the TESS fold in Figure~\ref{fig:rxjfig} shows four peaks,
the ground-based fold has two unequal peaks of higher amplitude.  TESS observed during a historical minimum state (see Section~\ref{sec:atlas}), while
the ground-based observations are concentrated in earlier years when the star was brighter.  The beat amplitude and profile may be sensitive to the average accretion rate in a way that reduces the amplitude and increases the complexity of the accretion flow when the accretion rate is lower.

The 29 individual optical light curves are shown in Figure~\ref{fig:phase_b},
folded on the Case~B spin period, which allows the simplest 
phase relationship.   The spin pulses come from either pole~1 or pole~2,
corresponding to peaks at integer or half-integer rotations, respectively.  In contrast, 
folding on the Case~A spin period (not shown) does not produce consistent phasing.
Peaks fall as far as 0.25~cycles from their expected location,
similar to the X-ray light curve when interpreted in Case~A.

On several nights when the star was faint, pulses were very weak, which
we attribute to a minimum of the accretion rate in the beat cycle.
On two nights is it possible to see pole switching taking place
at a beat minimum, where peaks come 0.5 cycles apart
(2016 March 17) or 1.5 cycles apart (2017 December 16).
The four most recent light curves,
in 2023 January, are simultaneous with the TESS observation,
which directly confirms their phasing, and the fact that the beat amplitude
is smaller than in 2014--2019.  The lack of data in the years 2020--2022 obscures any details of the transition.

\begin{deluxetable}{llcc}
\label{tab:mdm}
\tablecolumns{4} 
\tablewidth{0pt} 
\tablecaption{MDM/LCOGT Observed and Inferred Frequencies}
\tablehead{
 \colhead{Frequency} & \colhead{Period} & \colhead{Case A}
 & \colhead{Case B} \\
 \colhead{(day$^{-1}$)} & \colhead{} & & 
}
\startdata
& \hfill Observed & & \\
\hline
1.489498(12) &  16.11281(13) hr  & $\omega_{\rm A}-\Omega$  & $2(\omega_{\rm B}-\Omega)$ \\
14.63083(5)  &  98.42230(37) min & $\Omega$            & $\Omega$ \\
16.12028(8)  &  89.32847(44) min & $\omega_{\rm A}$            & $2\omega_{\rm B}-\Omega$ \\
\hline
&  \hfill Inferred  & & \\
\hline
0.744749(6) &  32.22562(26) hr  & \dots &  $\omega_{\rm B}-\Omega$ \\ 
15.37556(10) &  93.6552(6) min  & \dots &  $\omega_{\rm B}$ \\ 
\enddata
\end{deluxetable}

\begin{figure*}
\centerline{
\includegraphics[angle=270.,width=0.85\linewidth]{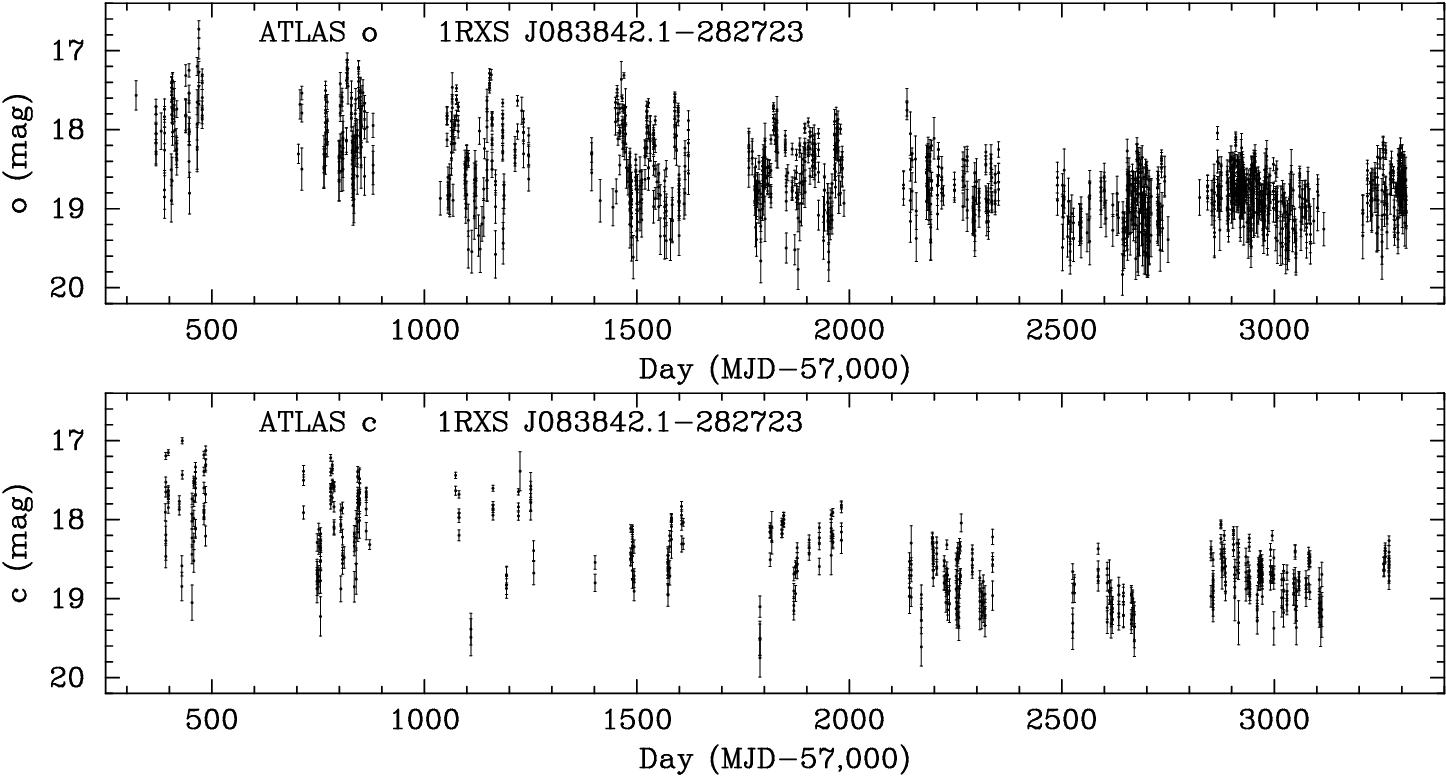}
}
\caption{
ATLAS photometry from 2015 October -- 2023 December in the $o$ and $c$ filters.
}
\label{fig:atlas}
\end{figure*}

\begin{figure*}
\centerline{
\includegraphics[angle=270.,width=0.8\linewidth]{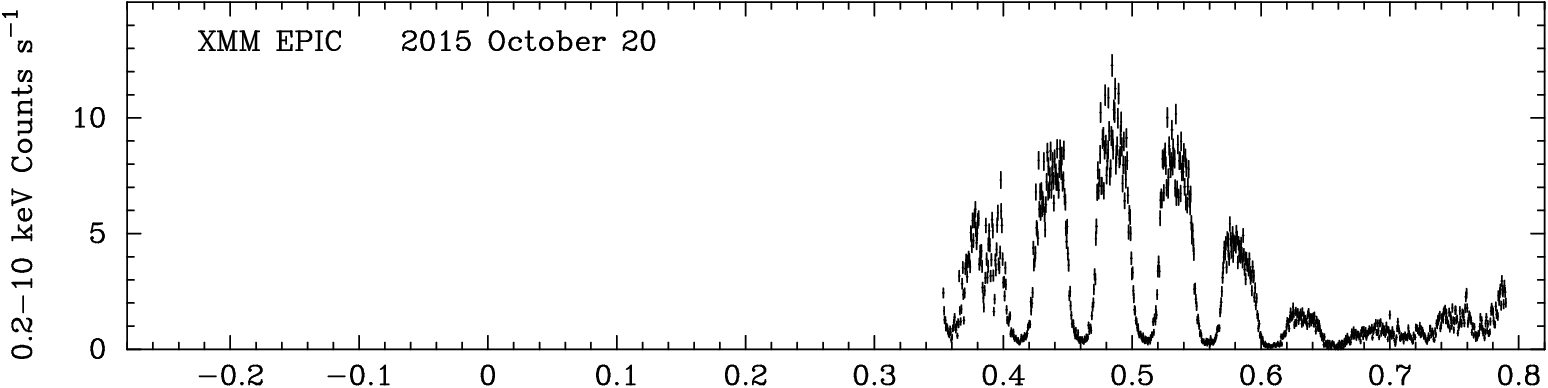}
}
\centerline{
\includegraphics[angle=270.,width=0.8\linewidth]{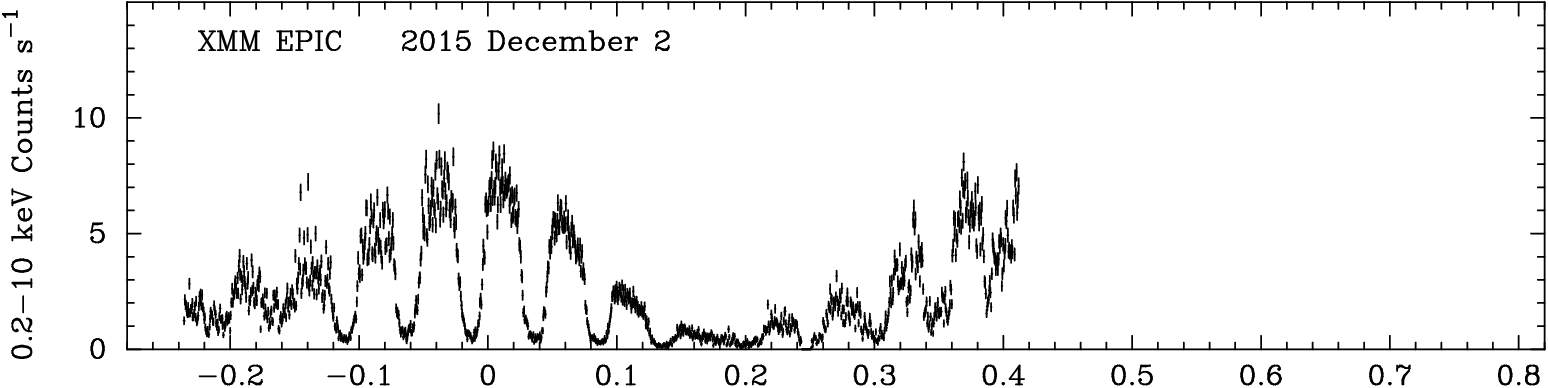}
}
\centerline{
\includegraphics[angle=270.,width=0.8\linewidth]{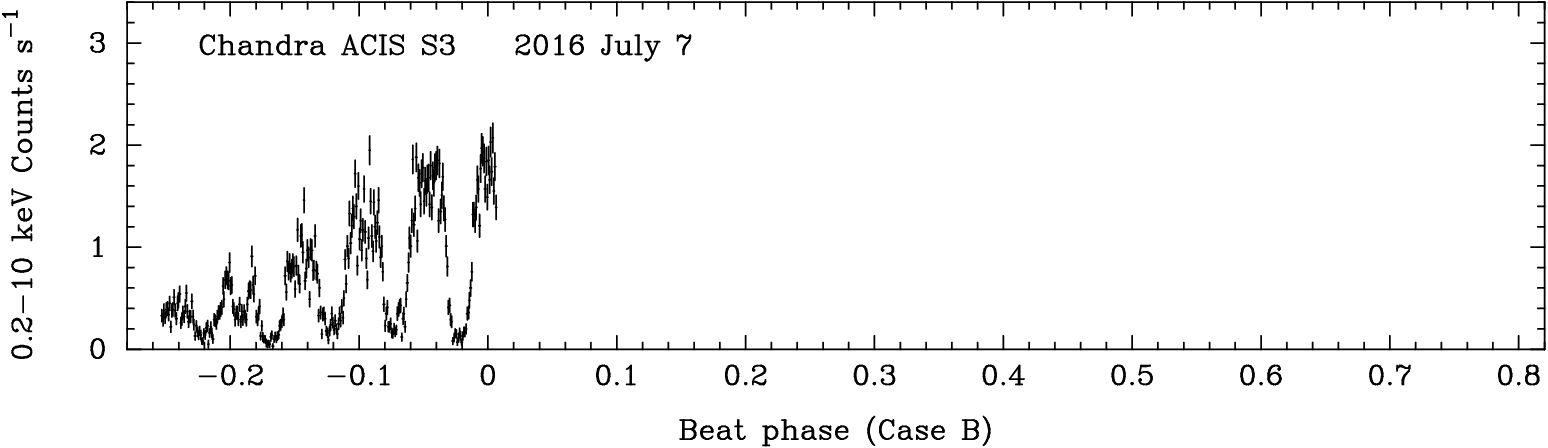}
}
\caption{All X-ray light curves of \rxj\ as a function of the Case~B beat phase defined in Table~\ref{tab:ephem}.
}
\label{fig:beat}
\end{figure*}

\subsection{ATLAS (2015--2023)\label{sec:atlas}}

We downloaded photometry of \rxj\ from the Asteroid Terrestrial-impact Last
Alert System (ATLAS; \citealt{ton18}) forced photometry server, which is
useful for documenting the long-term optical behavior of \rxj.  After
filtering out points with uncertainties $>0.3$~mag,
there are 1412 observations in the ``orange'' $o$ filter (560--820 nm)
and 543 observations in the ``cyan'' $c$ (420--650 m) filter
between 2015 October~26 and 2024 January~1. These data are shown
in Figure~\ref{fig:atlas}, where a slow decline of the mean magnitude
from $\approx 17.8$ to $\approx 19.0$ is seen in both filters.
Power-spectrum analysis of the barycentered data does not detect any significant periods, probably due
to the sparse sampling.  Nevertheless, it is evident that the day-to-day
scatter in the points has become smaller in recent years, likely
the effect of the decrease in the beat modulation seen in the TESS and
most recent MDM data.

\subsection{Constructing the Long-term Ephemerides\label{sec:ephem}}

Only by combining the X-ray, TESS, and ground-based timing can precise beat, orbit,
and pulse ephemerides be constructed that span 2014--2023.   The X-ray
observations were too far apart to count cycles between them
and obtain more precise periods.
Similarly, the TESS observation, even with its more precise periods, was too far
removed from the X-ray observations to perform that phase linkage.   However,
once the TESS periods were used to identify the more precise corresponding
values in the ground-based photometry, absolute phasing of all of the data follows.

All that remains is to define epochs of phase zero $T_0$ for each of the
periods. For the orbital period, we use the epoch of blue-to-red crossing
of the optical emission lines from Paper~1.
For the spin period, we use the time of a {\it minimum} in the \xmm\ light curve
of Figure~\ref{fig:rxjfig}.  Empirically, this is the most precise and least ambiguous fiducial,
representing the center of the self-eclipse of an emitting pole by the WD.
For the beat period, we estimate the time of {\it maximum} of the same X-ray
light curve.  The ephemerides so defined are given in Table~\ref{tab:ephem}.

\begin{figure*}
\vbox{\hspace{0.6in}
\hspace{-0.13in}
\includegraphics[angle=0.,width=0.8124\linewidth]{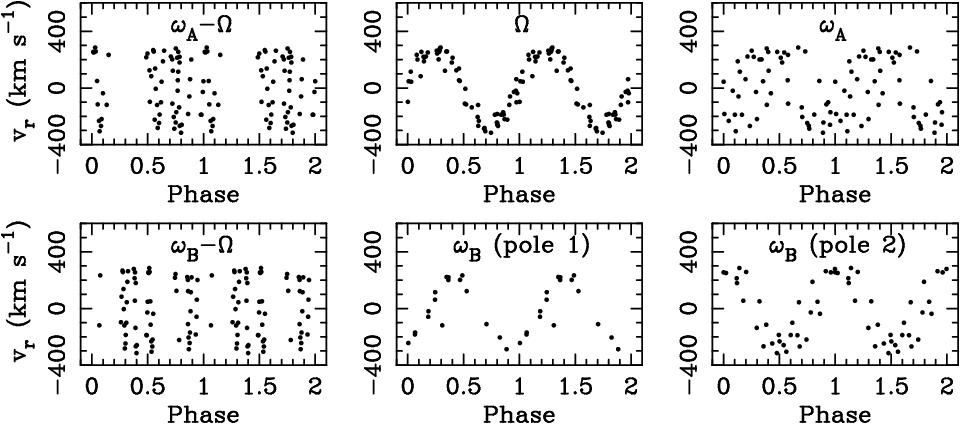}
}
\caption{
  Radial velocities of H$\alpha$ emission from Paper~1, folded on the beat,
  orbital, and spin ephemerides of Table~\ref{tab:ephem}.  Modulation
  is cleanest at the orbital period, but also strong at the Case~B
  spin period when taking into account pole switching.
}
\label{fig:halpha}
\end{figure*}

The phases of the folded light curves in Figures~\ref{fig:rxjfig}--\ref{fig:spin_a}
are all defined by these ephemerides.  They reveal several model-independent relationships:

\bigskip\noindent
1. The optical beat is in phase with the X-ray beat (Figures~\ref{fig:mdm}, \ref{fig:beat}).

\noindent
2. The X-ray observations together sampled all of the beat phases, even in
Case B (Figure~\ref{fig:beat}).

\noindent
3. The optical spin profile is in phase with the X-ray spin profile
(Figures~\ref{fig:rxjfig}, \ref{fig:phase_b}).

\noindent
4. With phase~0 as the epoch of blue-to-red crossing, the optical continuum brightness is in phase with the radial velocity curve of the optical emission lines (Figures~\ref{fig:rxjfig}, \ref{fig:mdm}, \ref{fig:halpha}).

\bigskip\noindent
The implications of these results will be discussed in Section~\ref{sec:disc}.

\begin{deluxetable}{lll}
\label{tab:ephem}
\tablecolumns{3} 
\tablewidth{0pt} 
\tablecaption{Ephemeris Parameters}
\tablehead{
\colhead{Quantity} & \colhead{$T_0$}  & \colhead{Frequency\tablenotemark{a}} \\
 & \colhead{(BJD)} & \colhead{(day$^{-1}$)}
}
\startdata
Orbit     & 2457408.8373(6)\tablenotemark{b}  & 14.63083(5)\\
\hline
& \hfill Case A  & \\
\hline
Spin & 2457359.741(1)\tablenotemark{c}& 16.12028(8)  \\
Beat & 2457359.76(1)\tablenotemark{d}  & 1.489498(12) \\
\hline
&  \hfill Case B   & \\
\hline
Spin & 2457359.741(1)\tablenotemark{c}  & 15.37556(10) \\
Beat & 2457359.76(1)\tablenotemark{d}  & 0.744749(6) \\
\enddata
\tablenotetext{a}{Ground-based frequencies listed in Table~\ref{tab:mdm}.}
\tablenotetext{b}{Epoch of blue-to-red crossing of radial velocities from
Paper~1, corrected to TDB.}
\tablenotetext{c}{Minimum of X-ray spin pulse profile in Figure~\ref{fig:rxjfig}.}
\tablenotetext{d}{Maximum of X-ray beat profile in Figure~\ref{fig:rxjfig}.}
\end{deluxetable}

\subsection{Phasing of Emission-Line Radial Velocities (2016)}

Using the precise ephemerides of Table~\ref{tab:ephem}, we graph the
59 emission-line radial velocities from Paper~1, obtained in 2016
January--February, as a function of beat,
orbit, and spin phase in Figure~\ref{fig:halpha}.  As expected,
there is no dependence on beat phase.
The sinusoidal modulation that was attributed to the orbital period in
Paper~1 is reproduced.  But now we also fold at the candidate spin
periods.  In Case~B this requires separating the points into two groups from different
beat phases in order to search for effects of pole switching, as the TESS data were treated in Section~\ref{sec:tess}.

In fact, modulation at the Case~B spin period is almost as clear as it is on the orbital period, once the two halves of the beat cycle
are separated, revealing spin cycles $\approx180^{\circ}$ out of phase.
Folding at the Case~A spin period also shows significant modulation,
but here the scatter is greater than in the Case~B curves.  The
scatter is similar in magnitude to the drift in the X-ray and optical
continuum pulses relative to the Case~A spin ephemeris, which is discussed in Section~\ref{sec:spin}.  Figure~\ref{fig:halpha} therefore suggests some preference for Case~B over Case~A.

Paper~1 concluded that the line-emitting region is fixed in the orbital frame,
probably in the ballistic accretion stream.  Now it seems equally possible
that the line emission is coming from high in the magnetically confined accretion
column, but fixed in the spinning frame.  For pole~1,
the maximum H$\alpha$ blueshift occurs at phase 0, when pole~1 is eclipsed,
while for pole~2 the eclipse comes half a spin period later.  Therefore,
the velocities are all consistent with radial infall. The orbital
longitude of the companion is not determined by the emission-line radial
velocity curve, but it is likely that maximum blueshift at phase $\approx0.75$
in the $\Omega$ panel of Figure~\ref{fig:halpha} corresponds to superior
conjunction, with the accretion flow being nearly radial.  With the reference
frame of the emission lines being ambiguous, it is possible that both orbit
and spin contribute.

\section{Discussion\label{sec:disc}}

\subsection{Orbital Period\label{sec:orbit}}

The goal of this study is to determine the true spin, orbit, and beat periods
of \rxj, none of which was securely known.  Our identification of the orbital
period is supported by consistency between the emission-line radial velocity period and an identical
optical photometric period.  However, it is not obvious in the case of an AP whether the optical emission lines {\it should\/} be
tied to the orbital frame of the secondary or the spinning frame of the WD.
It would be logical to assume the latter, that the strong
magnetic field captures the ballistic stream high up and confines the
emission-line region to the spinning frame.

Nevertheless, it is sometimes concluded that the emission-line radial velocity period is orbital when a different period in X-rays is judged more likely to be the spin, e.g., in \polb\ \citep{tov17} and \swj\ \citep{hal22}.  In CD~Ind and SDSS J134441.8+204408, \citet{lit19,lit23b} argued for a reinterpretation of the optical photometric period structure
in TESS data that would allow the spectroscopic period, previously assumed to be orbital, to be the spin period. \rxj\ presents a different problem in that pole switching in the radial velocities makes them identifiable to some extent with either frame, at least within the limited sampling of the existing spectroscopy.  Tentatively, both orbit and spin frames could be represented in the emission lines.  If photospheric absorption lines
can be detected from the companion, perhaps in the new low state of accretion, at least its orbital period and epoch of ascending node could be securely determined.

The radial-velocity amplitude of H$\alpha$ emission in
\rxj\ is $274\pm16$~km~s$^{-1}$ (Paper~1),
larger than that of most IPs, but similar to that of polars.  \cite{rea17}
decomposed higher-resolution emission-line profiles into three components,
one with an even larger $\approx1200$~km~s$^{-1}$ velocity amplitude that
they attributed to the ballistic accretion stream.
Both studies found that the equivalent width of the Balmer lines is higher
(by a factor of $\simle2$)
when their radial velocity is maximally blueshifted.  In Paper~1, we interpreted
this to mean that the ballistic accretion stream is maximally blueshifted when
the secondary is at superior conjunction, at the same time that the
optical continuum, produced lower in the accretion stream, is maximally occulted.  TESS confirms that the optical
continuum folded on our orbital frequency $\Omega$ is faintest at phase 0.75
(Figure~\ref{fig:rxjfig}), when the emission-line radial velocity
is maximally blueshifted (Figure~\ref{fig:halpha}).

\subsection{Beat Period\label{sec:beat}}

The beat period measured optically allows the X-ray light curves to
be aligned in absolute beat phase in Figure~\ref{fig:beat}, which
we illustrate in Case~B for evaluation of the ambiguity with Case~A.
If Case~B applies, then both halves of the beat cycle were sampled by
the combined \xmm\ observations taken 43 days apart.  But it is not
clear from these light curves that the second half of the cycle,
peaking at phase 0.5, differs in any way
from the first half, peaking at phase 0.
The similarity tends to argue instead for Case~A.

The spin pulses have basically the same shape in X-rays during all parts of the beat cycle, with modulation of $\approx100\%$.  Their absolute flux changed modestly between the two \xmm\ observations, but this may
represent long-term variability of the accretion rate rather
than a difference in the projected appearance of one pole versus the other.
The longest light curve, from 2015 December~2,
did not cover enough of the second half of the beat cycle
to prove that the second half differs from the first half in detail.
Thus, the evidence from X-rays alone does not strongly support Case~B.
At best, it requires a highly inclined and symmetric magnetic geometry
to account for the strong spin modulation and similar pulse profile from  both poles in Case~B, perhaps with both poles in the same (upper) hemisphere.
In addition, the accretion rate onto the WD has to go to zero during the switch between
the poles, with matter perhaps stored in the accretion ring
while the accretion stream is disconnected from either magnetic funnel.

\begin{figure*}[t]
\centerline{
\includegraphics[angle=270.,width=0.8\linewidth]{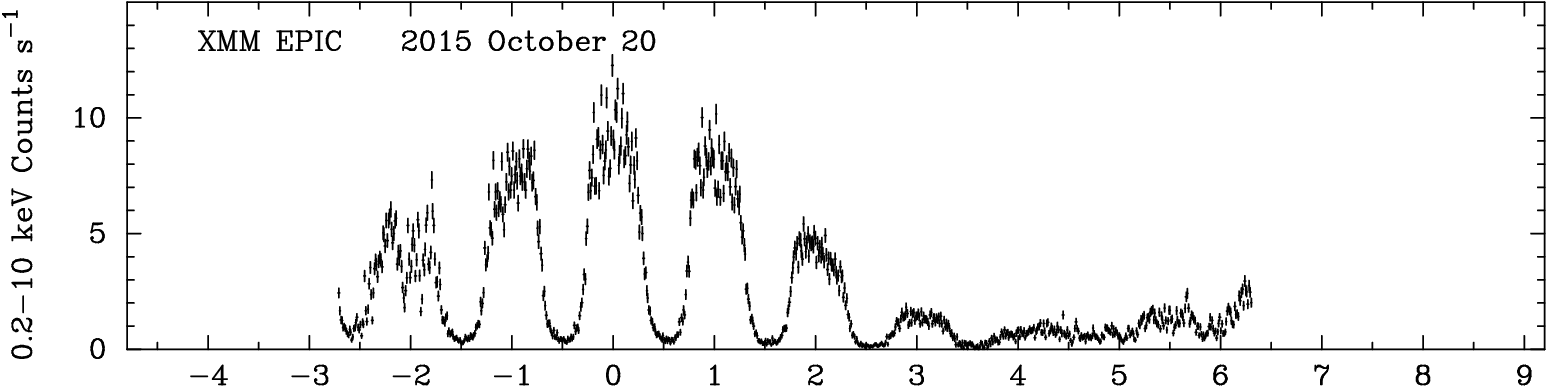}
}
\centerline{
\includegraphics[angle=270.,width=0.8\linewidth]{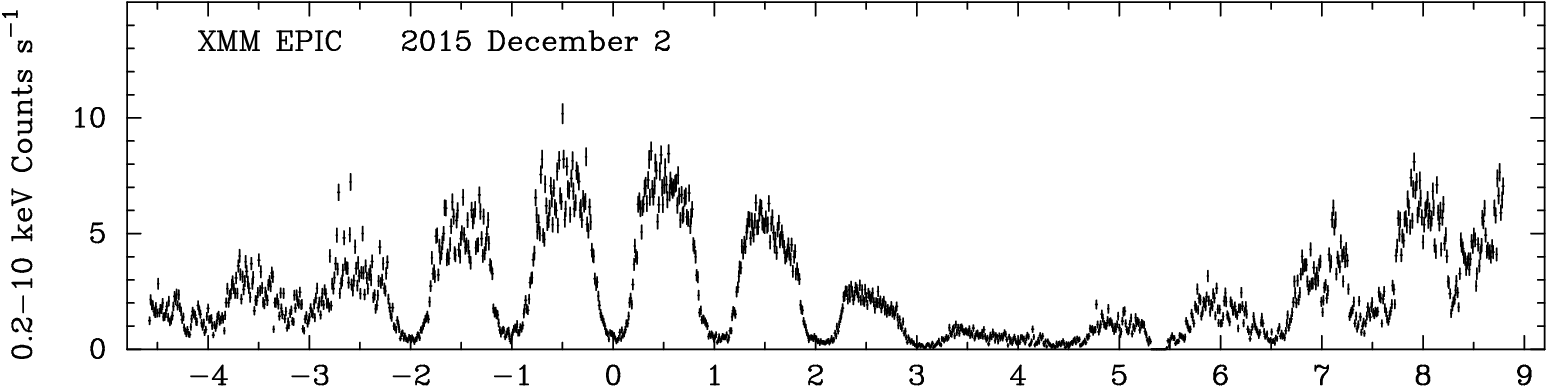}
}
\centerline{
\includegraphics[angle=270.,width=0.8\linewidth]{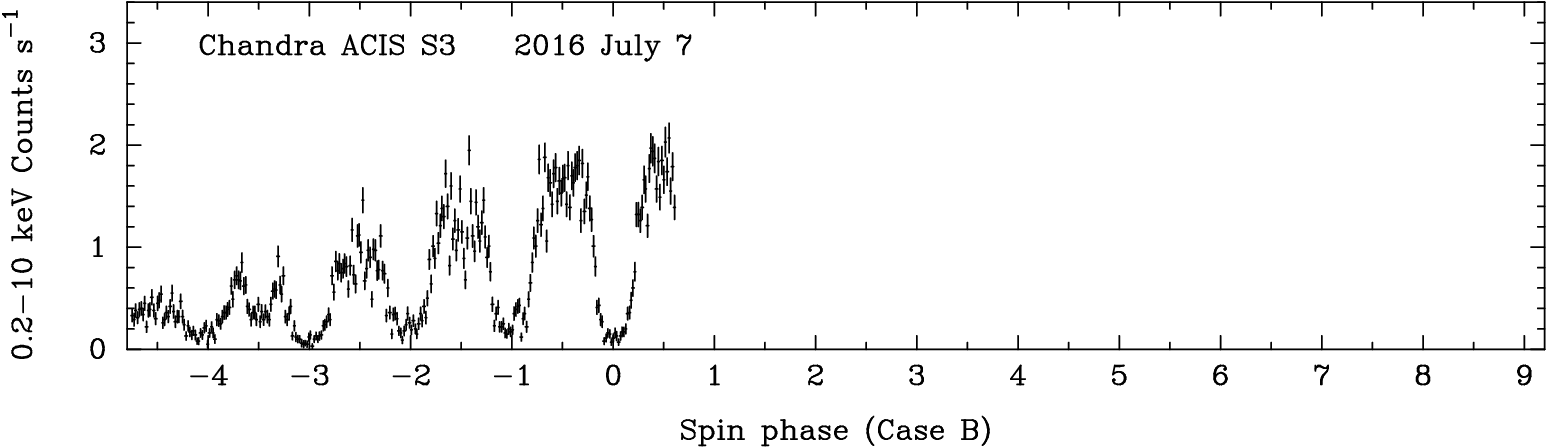}
}
\caption{All X-ray light curves of \rxj\ as a function of the Case~B spin phase defined in Table~\ref{tab:ephem}.
}
\label{fig:spin_b}
\end{figure*}

In Case~A, the beat modulation is caused by accretion switching from an
upper pole to a lower pole that is mostly invisible; thus, the X-ray flux can go to zero at minimum without the total accretion rate having to modulate. It is an effect of small magnetic and observer inclination angles in Case A. However, the spin period required in Case A faces other issues, which will be discussed in the next Section.

The optical observations are possibly more definitive of the beat period due to their long duration covering many beat cycles.
Analysis of the TESS data did isolate optical
emission at the Case~B spin period from two poles that differ in flux by
$\approx10\%$ averaged over the 50~days, and are aligned
in phase with the X-ray pulses (Figure~\ref{fig:rxjfig}).
This optical flux difference, also
manifest as the fundamental and multiples of the Case~B beat frequency
in the TESS power spectrum, appear to require Case~B. 
The two unequal peaks in the ground-based beat profile (Figure~\ref{fig:mdm}) also favor Case~B.

The multiple peaks in the TESS beat profile also suggests that the accretion rate was modulated in a more complex fashion in 2023, when
the optical emission was low, than it was in
the \xmm\ observations of 2015 (which included the Optical Monitor)
and the early ground-based
observations, when the optical emission was brighter.  It would now be
most interesting to obtain a long X-ray observation in a 
low optical state.  Since the potential locations of X-ray emission are more limited than the optical ones, the change in the accretion structure may be easier to diagnose in the X-ray.

\begin{figure*}[t]
\centerline{
\includegraphics[angle=270.,width=0.8\linewidth]{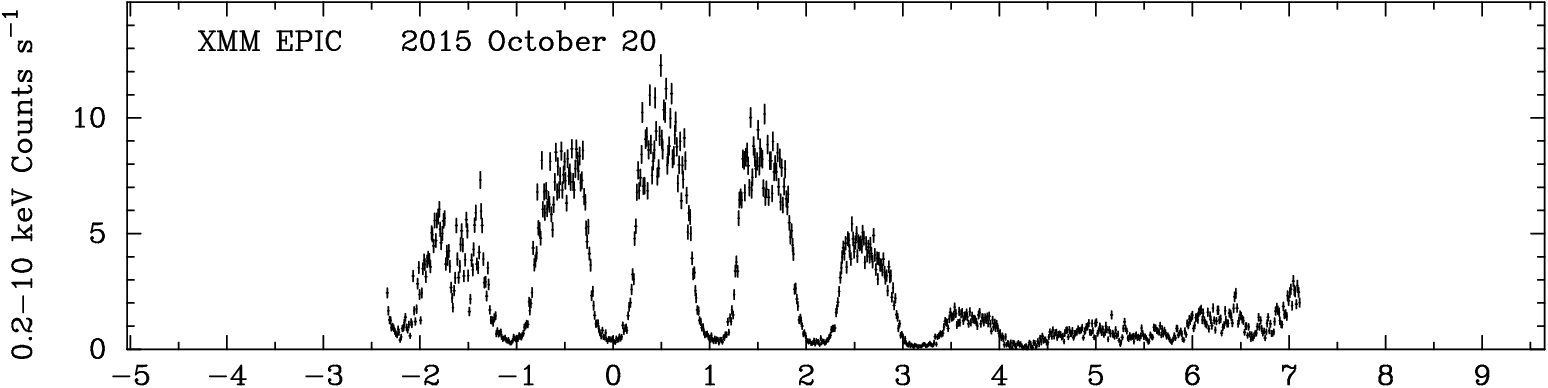}
}
\centerline{
\includegraphics[angle=270.,width=0.8\linewidth]{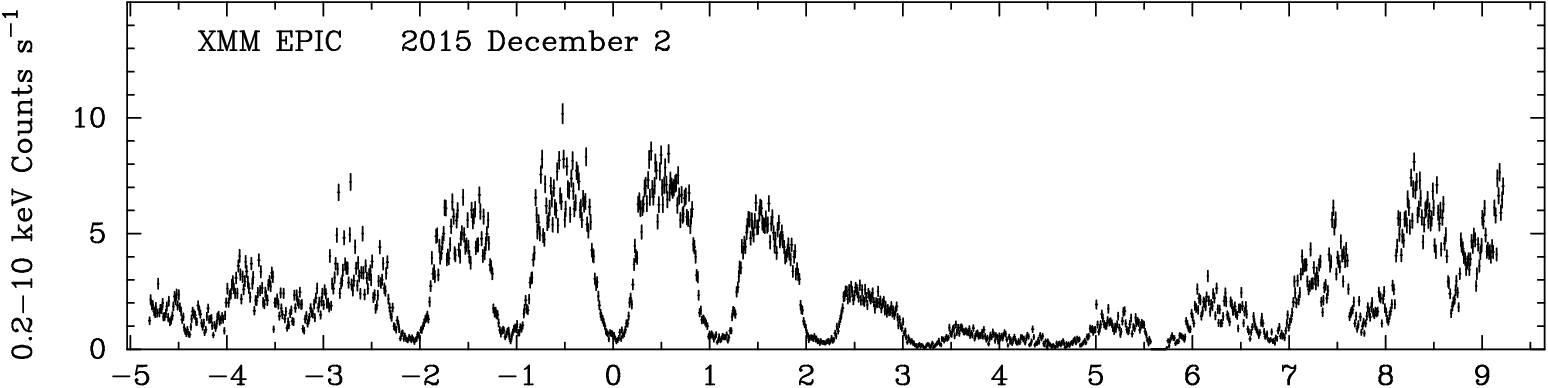}
}
\centerline{
  \includegraphics[angle=270.,width=0.8\linewidth]{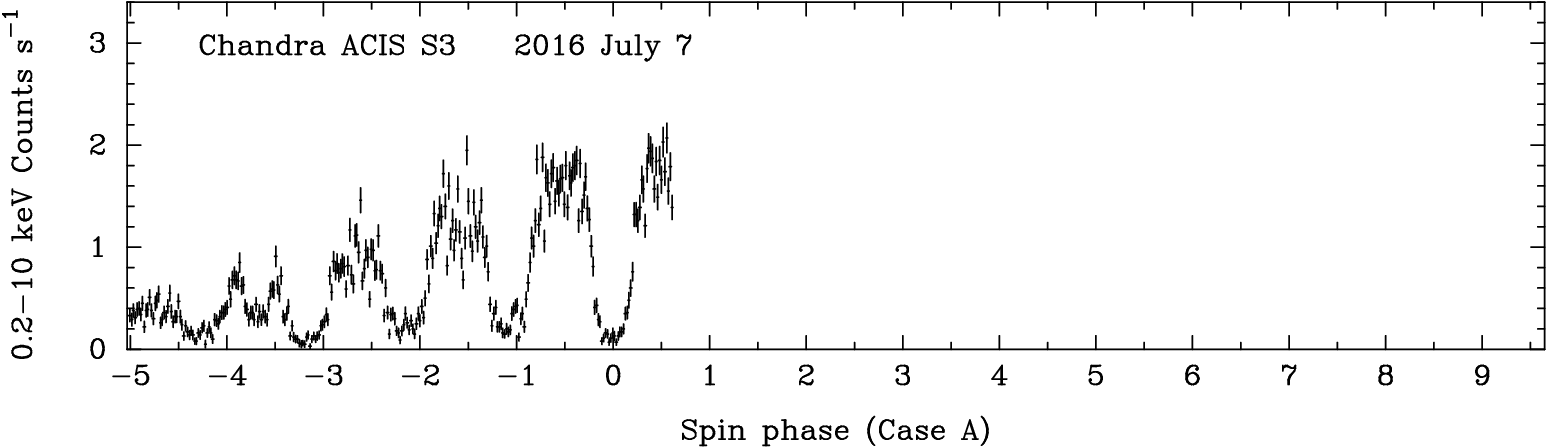}
}
\caption{All X-ray light curves of \rxj\ as a function of the Case~A spin phase defined in Table~\ref{tab:ephem}. 
}
\label{fig:spin_a}
\end{figure*}

\subsection{Spin Period\label{sec:spin}}

The fast X-ray modulation is most naturally attributed to the spin period
via self-occultation by the rotating WD of a shocked emitting region at
the base of the accretion column.  The observed dips are too broad to be
due to eclipse by the orbiting secondary, or by the accretion stream.
The value of the spin period follows directly
from the beat period, or vice versa, because a single accreting
pole is always dominant in the light curve in Case~A,
while two visible poles alternate during a twice
longer beat period in Case~B.

The period that most closely follows the X-ray and optical dips and peaks is the one that we infer is the spin period in Case~B.  This can be seen most easily in Figure~\ref{fig:phase_b} for the optical and Figure~\ref{fig:spin_b} for the X-ray.  But this works only if there is complete pole switching, such that the spin phase jumps by half a spin cycle at the transition between each half of the beat cycle.

In contrast, neither the assumed orbital period nor the Case~A spin period
coincides with the phasing of the optical or X-ray modulation.  This is
seen in Figure~\ref{fig:spin_a}, which tests the Case~A spin period
against the X-rays.  Here the peaks
drift by as much as $\pm90^{\circ}$ in phase during each beat cycle.  The
same effect occurs in the ground-based optical data (the Case~A version
of Figure~\ref{fig:phase_b}, not shown).

However, there is one caveat to this argument that may
lend some support to Case~A.  As first argued by \citet{gec97}, the footpoint of the accretion column on an AP is not expected to remain fixed on the surface of a WD.
As the secondary transits over a magnetic pole, the accreting matter is captured and threaded onto successively different field lines.  Looking from a fixed point on the WD, the secondary orbits in the retrograde direction if $\omega>\Omega$ as we infer here.
The effect on the pulse timing is to shift it to earlier times
at the beginning of the beat cycle, and to later times at the end of the cycle, with respect to the mean spin ephemeris.

This is in fact the sense in which the phase drifts in the Case~A
interpretation (Figure~\ref{fig:spin_a}) of \rxj. Such a spin-phase
drift is also seen in other APs \citep{lit19,lit23a},
although it is not clear if the model of \citet{gec97} quantitatively
explains the results.  The shifts are predicted to be of order $\pm20^{\circ}$
in longitude.  For \rxj\ in Case~A, the drift of $\pm90^{\circ}$ would imply
that the accretion column travels half the circumference of the WD.
That is, the footpoint of accretion approximately tracks the longitude of the companion.  This would be an interesting challenge for
modeling to address.

\section{Conclusions\label{sec:conc}}

\rxj\ has perhaps the simplest X-ray light curve of all APs, suggesting that the identification of its spin, orbital, and beat periods should be straightforward.  We used TESS and ground-based optical timing to construct precise ephemerides for all of these periods, phased to the X-rays, as explained in the footnotes to Table~\ref{tab:ephem}.  The original factor of 2 ambiguity in the beat period from the X-ray light curves appears to be resolved in favor of the longer beat period (Case~B) by structure in the optical beat profile, and complete pole switching between each half of the beat cycle as seen in the spin profile.  It requires the spin frequency $\omega_{\rm B}$ to be absent in the power spectrum, replaced by $\Omega$ and $2\omega_{\rm B}-\Omega$.  In this case $P_{\rm beat}=32.2256$~hr, $P_{\rm orb}=98.4223$~min, and $P_{\rm spin}/P_{\rm orb} = 0.95156$. 

This conclusion is not obvious from the original X-ray observations. While it is possible that a longer X-ray observation would have revealed differences between adjacent halves of the beat cycle, they may be subtle, e.g., requiring there to be opposite poles at similar latitude in one equatorial hemisphere of the WD.

There is one important difference between the cases that is obvious in both X-ray and optical.  Case~B allows the footpoints of the accretion columns to remain relatively fixed on the surface of the WD, being phased well to the spin period.
But if Case~A is correct, the spin period is present in the power spectrum, and being shorter than in Case~B, the footpoint of the single visible pole drifts in longitude almost halfway around the WD during each beat cycle.  The plausibility of either of these dynamics should be tested by physical modeling.

An added complication is the long-term decline by $\approx1$~mag over the 9-year monitoring period, with its possible affect on the accretion rate and geometry.  The X-ray and optical light curves obtained at the beginning of this period have simpler beat structure than the recent TESS observation made near minimum light.  This could indicate a major change in the accretion flow, for which there is no contemporaneous X-ray comparison. 

During previous \xmm\ observations of \rxj\ the Optical Monitor was not configured to resolve the orbit and spin modulations. A new, simultaneous X-ray/optical observation by \xmm\ with its Optical Monitor in fast mode could reveal much about the accretion structure, including any state change.  If spanning at least 32~hours, such an observation may, finally, decide the true spin and beat periods.

\begin{acknowledgements}
We thank the referee for a thorough evaluation of the manuscript. 
\xmm\ is ESA science mission with instruments and contributions directly funded
by ESA Member States and NASA.  MDM Observatory
is operated by Dartmouth College, Columbia University, The Ohio State University,
Ohio University, and the University of Michigan.
A Sinistro imager on a CTIO 1m telescope of the LCOGT network was also used. Data collected by the
TESS mission were obtained from the MAST data archive at the Space Telescope
Science Institute (STScI).  The ATLAS science products were made possible
through the contributions of the University of Hawaii Institute for Astronomy,
the Queen's University Belfast, the Space Telescope Science Institute,
the South African Astronomical Observatory, and The Millennium Institute
of Astrophysics (MAS), Chile.   
\end{acknowledgements}
  
\facility{XMM, McGraw-Hill, Hiltner, LCOGT, TESS.}

\bibliography{ms.bbl}
\bibliographystyle{aasjournal}

\end{document}